# Growth of epitaxial tungsten oxide nanorods

M.Gillet*, R. Delamare, E. Gillet
UNIVERSITE D'AIX-MARSEILLE – L2MP-UMR CNRS 6137
Faculté des sciences et techniques – case 151
52 Avenue Escadrille Normandie Niemen, 13397 Marseille cedex 20, France

**Abstract**

A simple vapour deposition technique was used to prepare $WO_3$ one-dimensional nanostructures. $WO_3$ is sublimated at a relatively low temperature (550°C) in air at atmospheric pressure. The sublimated species are condensed on mica substrate at 500°C. Single crystalline are grow in epitaxy on the mica surface with a growth axis along [010] directions and (001) plane parallel to the substrate. A growth process is proposed in which the formation of a one-dimensional tetragonal tungsten bronze as precursor is the determinant factor.

**Keywords**

A1: Nanostructures, A2: Growth from vapor, B1: Nanomaterials, B1:Oxides.

**Introduction**

Tungsten oxide has been widely studied and used in applications such as catalysis [1] electrochromic devices [2, 3, 4] or gas sensors [5, 6, 7]. Tungsten oxide $WO_3$ is a n type semiconductor with interesting properties as sensing material and recently it has been shown that nanostructured thin films have superior sensitivity compared to those of bulk material [8, 9]: Various methods including chemical vapour deposition [10], electrochemical deposition [11], laser vaporisation [12, 13] have been used to prepare tungsten oxide thin films. In conventional $WO_3$ thin films with nanoscale size grains, the electrical conduction is mainly controlled by the free carrier transport across the grain boundaries. So the synthesis of monocristalline tungsten oxide as nanowires or nanorods is of great interest. In the past years

Corresponding author: **M. Gillet**, UNIVERSITE D'AIX-MARSEILLE – L2MP-UMR CNRS 6137
Faculté des sciences et techniques – case 151, 52 Avenue Escadrille Normandie Niemen, 13397 Marseille cedex 20, France
Tel : +33-4-91-28-83-72 ; fax : +33-4-91-28-87-72, E-mail address : marcel.gillet@l2mp.u-3mrs.fr

[14] produced micrometer scale tree-like structure by heating a tungsten foil, partly covered by SiO$_2$ in Ar atmosphere at 1600°C. Theses nanostructures were composed by monoclinic W$_{18}$O$_{49}$ nanoneedles and by WO$_3$ nanoparticles. Nanorods of several oxides including WO$_3$ have been prepared by templating on acid-treated carbon nanotubes [15]. By heating WS$_2$ in oxygen, fibers of W$_{18}$O$_{49}$ were produced with a pine-tree like structure [16]. Mixtures of WO$_2$ and WO$_3$ with nanorods structure were obtained by koltyptin et al [17] via amorphous tungsten oxide nanoparticles. Y.B. Li et al [18] have synthesized WO$_3$ nanobelts and nanorods via physical vapour deposition process where the nanostructures were deposited on silicium wafers maintained at 600°C. Recently Z. Liu et al [19] reported on the preparation of tungsten oxide nanowires through a vapour-solid growth process by heating a tungsten wire partially wrapped with boron oxide at 1200°C and Y. Shingaya et al. [20] prepared by oxidation at high temperature well oriented WO$_x$ nanorods on a (001) W surface.

In this paper we report on the formation of tungsten oxide by a simple method using a vapour-solid growth process. Tungsten oxide is sublimated from a predeposited WO$_3$ layer and condensed on a mica substrate. A growth process is proposed where the formation of a tetragonal tungsten bronze acts as a precursor for the epitaxial growth of the nanorods.

**Experimental procedure**

Figure 1 shows the experimental set up that we have used to produce tungsten oxide nanorods. The tungsten oxide vapour source is a tungsten oxide thin layer predeposited on a SiO$_2$ substrate heated at a temperature T$_1$ by an electrical heater. The WO$_3$ vapour condenses on a substrate located above the vapour source at a distance d of the source by means of a wedge made in silicon. The temperature T$_1$ and the distance d determine the substrate temperature T$_2$. The experiments were conducted in a chamber where the humidity was controlled. In the our experiments $T_1$ was fixed to 550°C, and the resulting value of the substrate temperature $T_2$ was 450° ± 10°C.

We have used (0001) mica, (0001) $Al_2O_3$ and $SiO_2$ surfaces as substrates. Depositions were performed at atmospheric pressure in air with a degree of humidity comprise between 30 and 40%. After cooling at room temperature the substrate surface above the source had a faint yellow colour. The deposits were examined on their substrates by Atomic Force Microscopy (AFM). The structure of nanorods was investigated by High Resolution Transmission Electron Microscopy (HRTEM) and Transmission Electron Diffraction (TED). For electron microscopy observations a carbon replica was deposited on the sample surfaces and stripped off in water. The transfer replica contained a number of nanorods well suitable for TEM and TED investigations.

**Results**

The AFM observations showed that tungsten oxide nanorods grown on a mica substrate. On $Al_2O_3$ and $SiO_2$ substrates, only three-dimensional aggregates were observed. Experiments in air with various degrees of humidity have shown that $WO_3$ nanorods can be obtained in air with humidity in a range of 10-40%. Figure 2a is a typical AFM image obtained of a mica substrate after 45 minutes deposition time. It shows nanorods grown along the substrate with well defined directions. One or two nanorods orientations are predominantly observed. The tungsten nanorods size depends on the deposition time but for a given deposition time the size is not uniform. However the length seems to be independent of the deposition time while the width and the thickness increase with as illustrated by the figure 2b corresponding to a deposition time of 60 min. In our experiments the deposition time varied between 10 and 90 mn. The nanorod dimensions vary in ranges 1-15µm, 10-200nm and 1-50nm for the length, width and thickness, respectively. The observed nanostructures often exhibit multiple rods with twinned parts. Figure 3a shows as example such a double rod with one part grown in twin position. As the deposition time increases multitwining occurs as shown on figure 3b.

As mentioned in the experimental procedure, for electron microscopy investigations, the nanorods are extracted from their substrate by a transfer replica and observed without the mica substrates. The Energy Dispersive Xray Spectroscopy (EDX) shows that the nanorods contain potassium atoms (figure 4). Whatever their thickness the observed nanorods have approximatively the same amount of potassium as deduced from their EDX spectrum. We suppose that the nanorods grown on mica are initially composed of tungsten bronze ($K_xWO_3$)

Figure 5 and 7 illustrate two types of ED patterns from nanorods with two different thicknesses. The nanorod of figure 5 is a typical ED pattern of a thin nanorod e $\cong$3nm with a rectangular basic cell the interatomic distances as deduced from the ED are d1=0.627nm d2=0.382nm. This pattern can be interpreted as due to an hexagonal tungsten bronze (HTB) with lattice dimensions a=b=6.25 Å and c= 3.83 Å. From the ED we deduced that the HTB crystals grown on the mica surface with a (100) plane parallel to the substrate. This HTB structure is illustrated by the HRTEM image of figure 6 corresponding to the ED of the figure 5. The enlarged part inserted in the figure 6 shows the rectangular unit mesh and indicates the atomic distance in a (100) plane parallel to the surface of the nanorod, the second kind of ED patterns observed on the nanorods is represented on the figure 7 which exhibits two kinds of diffraction spots: Spots with strong intensities correspond to lattice distances $d_{200}$=0.367 nm Å and $d_{020}$= 0.378 nm which are very close of the theoretical distance in bulk $WO_3$ ($d^{th}_{200}$= 3.653 and $d^{th}_{020}$=3.77) and spots with low intensities located at half the distance between the bright spots. We interpret this electron diffraction pattern by the coincidence of two phases: a $WO_3$ phase with a monoclinic structure (a= 7.29 Å, b= 7.53Å, c= 7.68Å, β=90.91°) and a second phase with doubled parameters a and b. We deduce that the nanorod has a $WO_3$ monoclinic structure with [001] axis perpendicular to the substrate and a growth direction parallel to the [010] $WO_3$ direction. The figure 8 shows a HRTEM image corresponding to the ED of the figure 7. It exhibits a nearly square unit mesh characteristic of a (001) plane of the $WO_3$ monoclinic structure. The dimensions of the unit mesh as measured on the HRTEM are indicated on the enlarged parts inserted in the figure 8. According to these results we consider that in the first stage of growth, hexagonal tungsten bronze nanostructures epitaxially grow on the mica substrate and further growth lead to $WO_3$ nanorods with a monoclinic structure.

**Growth mechanism**

The tungsten oxide nanorods grow from deposition of vapour containing $WO_3$ generated at relatively low temperature, in air with a humidity degree which varies in a 20-40% range. Nanorods growth only takes place on mica substrate and in well defined orientations on the substrate. We propose a growth process based on these experimental results. It is known that many whiskers and nanowires are grown by a vapour-liquid-solid mechanism [21] in general perpendicular to the substrate and have a droplet on their tip [22]. In our experiments we did not observe droplet on the end of tungsten oxide nanorods, this indicates that the growth proceeds by a vapour-solid (VS) growth mechanism.

Experimental results show that the $WO_3$ nanorods are grown from $WO_3$ species sublimated in air containing water vapour, considerably lower than the sublimation temperature of bulk $WO_3$ (1470°C). We cannot exclude that the water vapour plays a role in the sublimation process, according to a reaction:

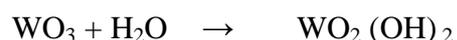

$$WO_3 + H_2O \rightarrow WO_2(OH)_2$$

On an other hand the nanorods do not grow on $Al_2O_3$ or $SiO_2$ substrates but only on mica substrates containing potassium ions in addition to the chemical elements present in $Al_2O_3$ and $SiO_2$. One deduces from the observations that in this special case the tungsten oxide growth is driven by the interaction between the species of the mica substrate and the deposit to form tungsten bronze $K_xWO_3$. Due to the role of the water vapour the following reaction:

$$WO_2(OH)_2 + x\,K \rightarrow K_xWO_3 + H_2O$$ seems the most probable.

Similar processes involving $WO_2(OH)_2$ species have been proposed in the growth mechanism of $K_{0.4}WO_3$ whiskers via an hydrothermal process using $WO_3$ as tungsten source, KOH as potassium source and hydrazine hydrate as reducing agent [23].

In the potassium tungsten oxide bronze ($K_xWO_3$), the W atoms are octahedrally coordinated with oxygen atoms to form $WO_6$ octaedrons. The $WO_6$ octaedrons join each other by shearing oxygen corner atoms. Tungsten bronzes can adopt various types of structures depending on ionic radius of the incorporated metal atom and on the composition. Hexagonal tungsten bronzes (HTB) can be formed with K ions in the composition range $0.13<x<0.33$. In the HTB the $WO_6$ octaedrons form hexagonal tunnels where the potassium ions can be located.

So we suppose that an epitaxial HTB grow on the surface of the mica substrate and acts as a precursor for the further $WO_3$ nanorod growth. The resulting monoclinic $WO_3$ structure result either of the growth of $WO_3$ on the HTB or of the growth of an hexagonal $WO_3$ followed by the transition of the hexagonal structure to a monoclinic structure. Hexagonal tungsten oxide is metastable phase and transforms irreversibly into the monoclinic structure $WO_3$. This transformation has been studied in detail by M. Figlarz et al. [24]. From the AFM observations it seems that the $WO_3$ growth proceeds layer by layer increasing the thickness and the width of the nanorods. During the growth twinning occurs along the [010] direction of the nanorod. The two twinned nanostructures are tilted relative to each other by about $2\beta-180°$ (monoclinic angle $\beta = 90.91°$). During the deposition the twinning can be repeated resulting in a multi twinned nanorod as shown on AFM images (figure 3b) or on TEM image (figure 4b). Such twinning is a relative common feature in $WO_3$ material. It has been observed by LEED on monocristalline $WO_3$ [24] and in $WO_3$ grain in epitaxial $WO_3$ thin film by ED [25] resulting in diffraction spot splitting.

**Conclusion**

Tungsten oxide nanorods have been epitaxially grown on mica using a simple vapour solid growth process. Experiments on different substrates (Mica, $Al_2O_3$, and $SiO_2$) have shown that the presence of potassium atoms is determinant for the nanorods growth. We deduce a growth process involving the formation of a one dimensional tetragonal tungsten bronze epitaxially oriented on the mica. This tetragonal tungsten bronze is a precursor for the $WO_3$ nanorod growth. Its formation on the mica substrate determines the nanorod morphology and orientations.

**Acknowledgements**

The work is supported by the European contract "Nanostructures for chemical sensors" (Project of the sixth Framework Programme).

**References:**

[1] F.A. Cotton and G. Wilkinson In: Advances in Organic Chemistry (5th ed.), Wiley, New York, 1988, p.829.
[2] C.G. Granqvist. Sol. Energy Mat. Sol. Cells 60 (2000) 201.
[3] B.P. Jelle and G. Hagen. Sol. Energy Mat. Sol. Cells 58 (1999) 277.
[4] I. Turyan, U.O. Krasovec, B. Orel, T. Saraidorov, R. Reisfeld and D. Mandler. Adv. Mater. 12 (2000) 330.
[5] W.M. Qu and W. Wlodarski. Sens. Actuators, B 64 (2000) 42.
[6] K.H. Lee, Y.K. Fang, W.J. Lee, J.J. Ho, K.H. Chen and K.S. Liao. Sens. Actuators B 69 (2000) 96.
[7] E. Llobet, G. Molas, P. Molinas, J. Calderer, X. Vilanova, J. Brezmes, J.E. Sueiras and X. Correig. J. Electrochem. Soc.147 (2000) 776.
[8] J. Tamaki, A. Hayashi, Y. Yamamoto, M. Matsuoka, Sens. Actuators B 95 (2003) 111.
[9] M. Gillet, K. Aguir, M. Bendahan, P. Menneni, Accepted in Thin Solid Films.
[10] E. Brescacin, M. Basato and E. Tondello. Chem. Mater. 11 (1999) 314.
[11] Z.R. Yu, X.D. Jia, J.H. Du and J.Y. Zhang. Sol. Energy Mat. Sol. Cells 64 (2000) 55.
[12] M. Sun, N. Xu, J.W. Yao and E.C. Wang. J. Mater. Res. 15 (2000) 927.
[13] S.T. Li and M.S. El-Shall. Nanostruct. Mater. 12 (1999) 215.
[14] Y.Q. Zhu, W. Hu, W.K. Hsu, M. Terrones, N. Grobert, J.P. Hare, H.W. Kroto, D.R. M. Walton and H. Terrones, Chem. Phys. Letters 309 (1999) 327.
[15] B.C. Satishkumar, A. Govindaraj, M. Nath and C.N.R. Rao, J. Mater.Chem 10 (2000) 2115.
[16] W.B. Hu, Y.Q. Zhu, W.K. Hsu, B.H. Chang, M. Terrones, N. Grobert, H. Terrones, J.P. Hare, H.W. Kroto and D.R.M. Walton. Appl. Phys. A 70 (2000) 231.
[17] Y. Koltypin, S. I. Nikitenko and A. Gedanken, J. Mater. Chem. 12 (2002) 1107.
[18] Y.B. Li, Y. Bando, D. Golberg and K. Kurashina, Chem.Phys.Letters, 367 (2003) 214.
[19] Z. Liu, Y. Bando and C. Tang, Chem.Phys.Letters, 372 (2003) 179.
[20] R.S. Wagner and W.C. Ellis. Trans. Met. Soc. AIME 233 (1965) 1053.
[21] H. Yumoto, T. Sako, Y. Gotoh, K. Nishiyama and T. Kaneko, J. Cryst. Growth 203 (1999) 136.
[22] X. Yang, C. Li, M.S. Mo, J. Zhan, W. Yu, Y. Yan, and Y. Qian, J. Cryst. Growth 249 (2003) 594.
[23] R.J.D. Tilley, Int. J. of Refractory and Hard materials 13 (1995) 93.
[24] F.H. Jones, K. Rawlings, J.S. Foord, R.G. Egdell, J.B. Pethica, B.M.R. Wanklyn, S.C. Parker, P.M. Oliver, Surf.Sci.359 (1996) 107.
[25] M. Gillet, A. Al- Mohammad, C. Lemire, Thin Solid Films 410 (2002) 194.

**Captions:**

**Figure 1:** Experimental set-up of the $WO_3$ deposition process. $T_1$= substrate temperature, $T_2$=Temperature of the $WO_3$ source, d= distance between the source and the substrate. ·

**Figure 2:** (a) AFM image of $WO_3$ nanorods grown on mica. Deposition time: 45 mn.

(b) AFM image of $WO_3$ nanorods grown on mica. Deposition time: 60 mn.

**Figure 3:** (a) AFM image of twinned nanorods. In inset: thickness profile according to the cross section A

(b) AFM image of multi-twinned nanorods (Tridimensionnal view).

**Figure 4:** EDX spectrum of a $WO_3$ nanorod. The elements corresponding to the various peaks are indicated.

**Figure 5:** Electron diffraction pattern of a thin nanorod ( e$\cong$ 3nm).

**Figure 6:** HRTEM image corresponding to the ED pattern of the figure 5. The unit mesh is indicated on the enlarged part (inset).

**Figure 7:** Electron diffraction pattern of a $WO_3$ nanorod ( thickness: 5nm).

**Figure 8:** HRTEM image corresponding to the ED pattern of the figure 7. The unit mesh is indicated on the enlarged part (inset).

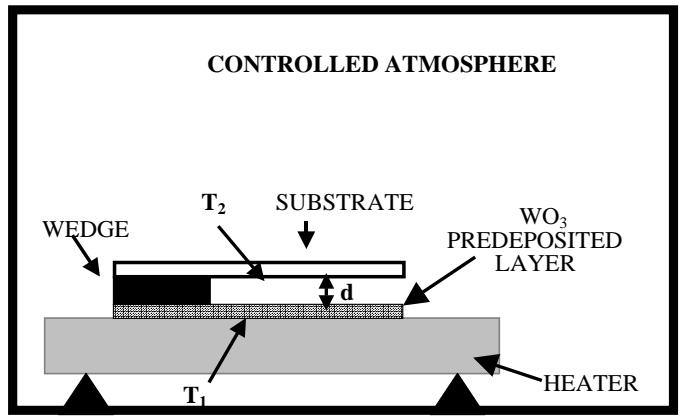

**Figure 1**

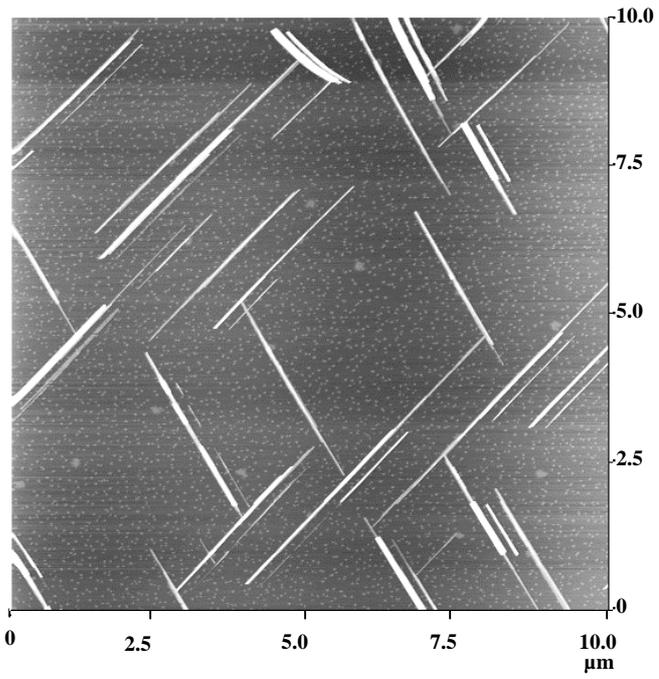 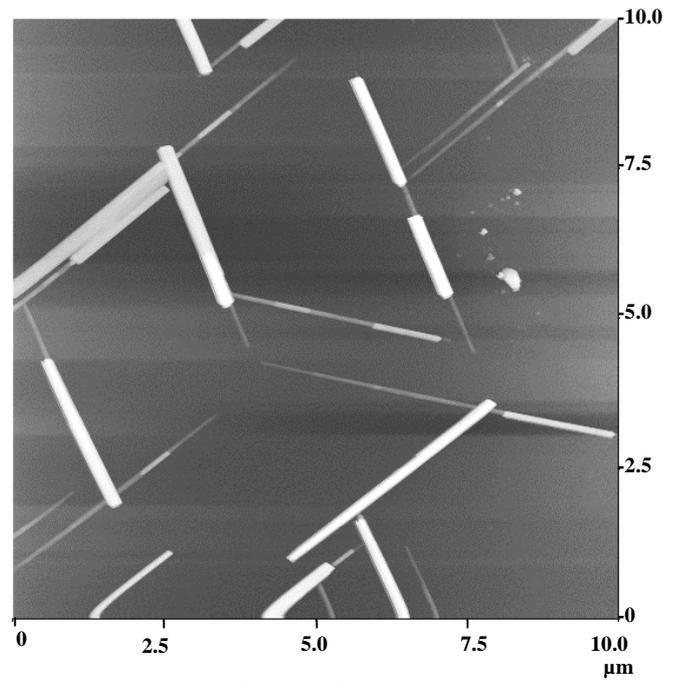

**Figure 2a** **Figure 2b**

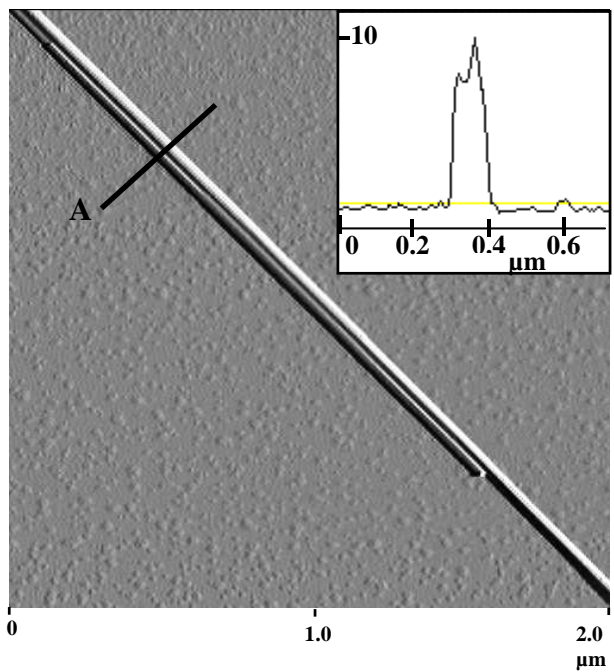

**Figure 3a**

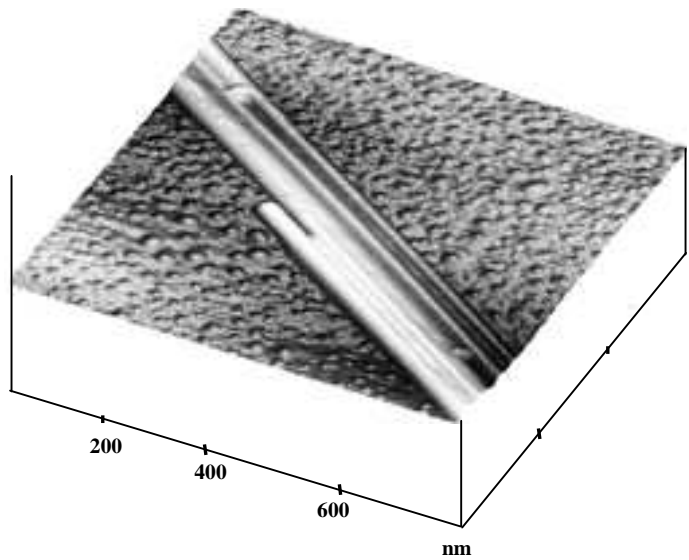

**Figure 3b**

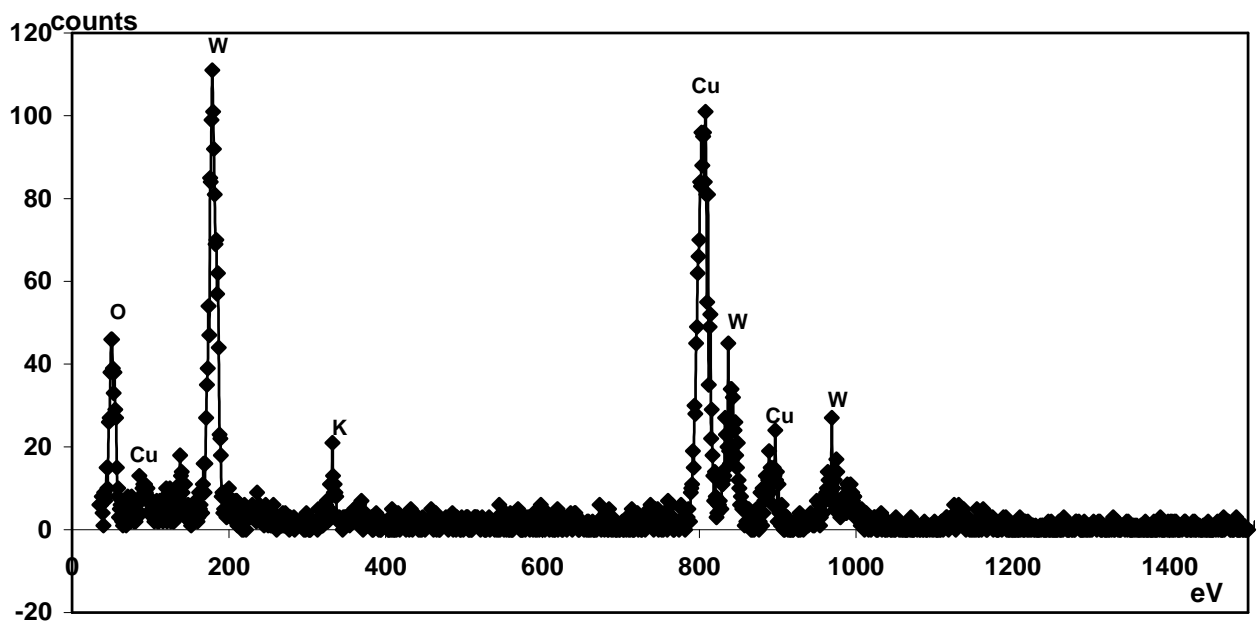

**Figure 4**

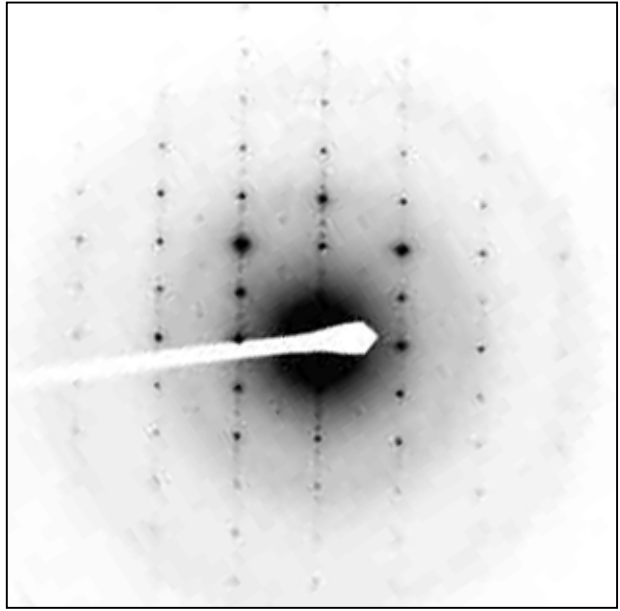

**Figure 5**

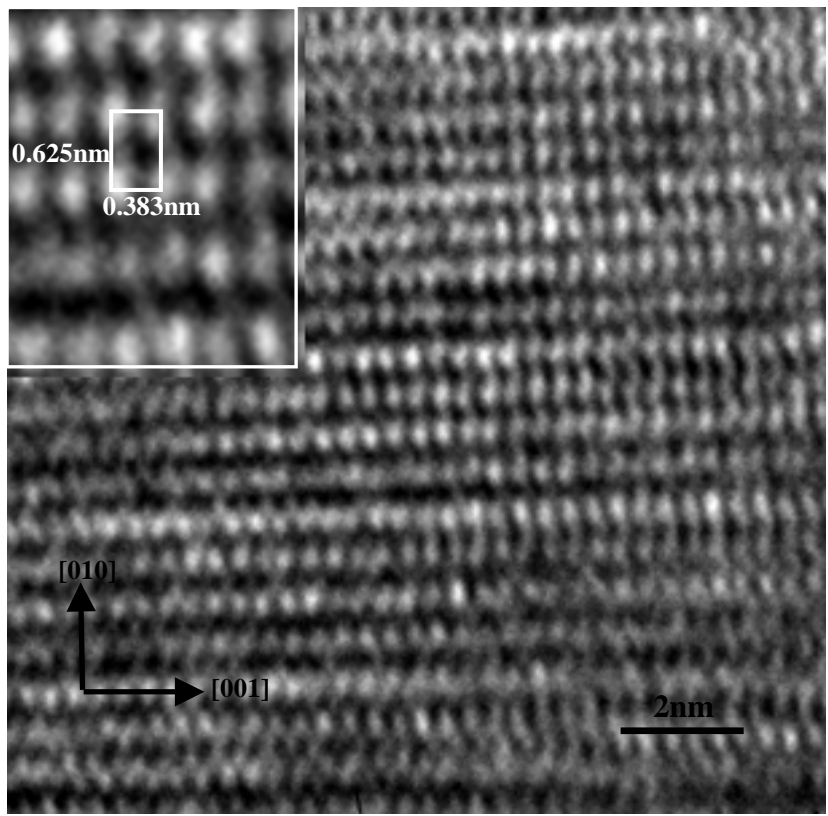

**Figure 6**

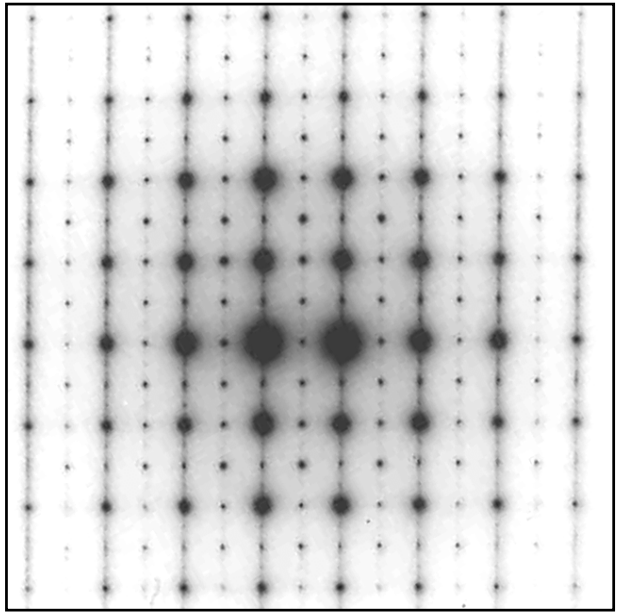

**Figure 7**

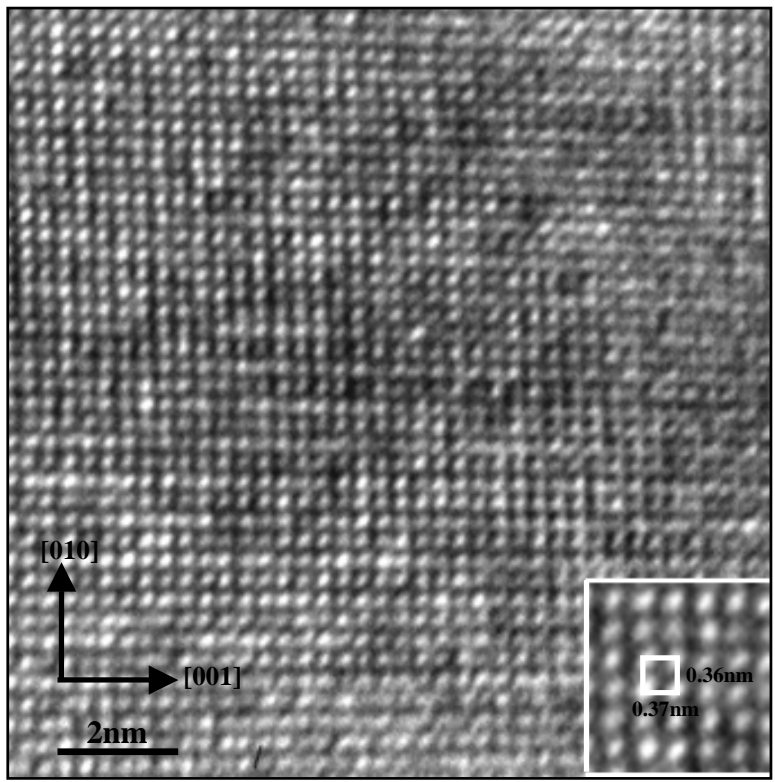

**Figure 8**